\documentclass{mem}
\usepackage{natbib}\usepackage{txfonts}\usepackage{balance}
\usepackage{graphicx}
\usepackage[a4paper,breaklinks,dvipdfm]{hyperref}
\idline{00}{000}
\begin{document}

\newcommand{\be}{\begin{equation}}
\newcommand{\ee}{\end{equation}}
\newcommand{\bea}{\begin{eqnarray}}
\newcommand{\eea}{\end{eqnarray}}

\title{
Standard and exotic singularities regularized by varying constants
}

   \subtitle{}

\author{
M.P. \,D\c{a}browski\inst{1,2}, K. \, Marosek\inst{1,3},
\and A. Balcerzak\inst{1,2}
          }

  \offprints{M.P. D\c{a}browski}

\institute{
Institute of Physics, University of Szczecin, Wielkopolska 15, 70-451 Szczecin, Poland
\and
Copernicus Center for Interdisciplinary Studies,
S{\l }awkowska 17, 31-016 Krak\'ow, Poland
\and
Chair of Physics, Maritime University, Wa{\l}y Chrobrego 1-2, 70-500 Szczecin, Poland
\email{mpdabfz@wmf.univ.szczecin.pl}
}

\authorrunning{D\c{a}browski}

\titlerunning{Varying constants regularized singularities}

\abstract{We review the variety of new singularities in homogeneous and isotropic FRW cosmology which differ from standard Big-Bang and Big-Crunch singularities and suggest how the nature of these singularities can be influenced by the varying fundamental constants.
}
\maketitle{}

\section{Introduction}

Currently, one is able to differentiate quite a number of cosmological singularities with completely different properties from a Big-Bang or a Big-Crunch. Many of them do not exhibit geodesic incompleteness, but they still lead to a blow-up of various physical quantities (scale factor, mass density, pressure, physical fields). In this paper we will discuss how they can be influenced by the variability of the fundamental constants.

\section{Standard and exotic singularities in cosmology.}

Standard Einstein-Friedmann equations are two equations for three unknown functions of time $a(t), p(t), \varrho(t)$ - the scale factor, the pressure, the mass density. In order to solve them, usually the equation of state is specified. Most common form of it is a barotropic one $p = w \varrho c^2$ with $w$ being a barotropic index, $c$ velocity of light. However, it is interesting that one obtains the independent evolution of the mass density and pressure provided we do not assume any equation of state which tights these quantities.

Until quite recently, including first supernovae results \citep{Perl99}, most cosmologists studied only the simplest - say ``standard'' solutions of the Friedmann equation. They each begin with a Big-Bang (BB) singularity for which $a \to 0$, $\varrho, p \to \infty$, while in the future one of them (of positive curvature $k=+1$) terminates at a second singularity (Big-Crunch - BC), where $a \to 0$, $\varrho, p \to \infty$ and the other two ($K=0, -1$) continue to an asymptotic emptiness $\varrho, p \to 0$ for $a \to \infty$. BB and BC exhibit geodesic incompletness and a curvature blow-up. In fact, the first supernovae observations gave evidence for the strong energy condition (SEC) ($\varrho c^2 + 3p \geq 0$, $\varrho c^2 + p \geq 0$) violation, but the paradigm of the ``standard'' Big-Bang/Crunch singularities remained untouched.

However, a combined bound on the dark energy barotropic index $w$ \citep{Tegmark2004} showed that there was no sharp cut-off of the data at $p = -\varrho$ so that
the dark energy with $p < -\varrho$ (phantom), could also be admitted. This lead to the cosmic ``no-hair'' theorem violation since even a small fraction of phantom dark energy could dominate the evolution instead of the cosmological term. Since $w<-1$ for phantom, then we may define $\mid w + 1 \mid = - (w+1) > 0$,
so $a(t) = t^{- 2/ 3 \mid w +1 \mid}$ and the conservation law gives $\varrho \propto a^{3 \mid w+1 \mid}$. This means that if the universe grows bigger, its density is higher, and finally it becomes dominated by phantom dark energy. An exotic future singularity -- a Big-Rip (BR) -- appears, for which $\varrho, p \to \infty$ for $a \to \infty$ \citep{phantom}. At Big-Rip the null energy condition (NEC), the weak energy condition (WEC), and the dominant energy condition (DEC) are all violated. Also, the curvature invariants $R^2$, $R_{\mu\nu}R^{\mu\nu}$, $R_{\mu\nu\rho\sigma}R^{\mu\nu\rho\sigma}$ diverge in the same way as at BB and at BC, and there is a geodesic incompletness at Big-Rip as well. Besides, everything is pulled apart on the approach to such a singularity in a reverse order.

Observational support for a Big-Rip gave an inspiration for studies of other exotic types of singularities as possible sources of dark energy. Especially, \citet{SFS} showed that if one drops an assumption about the imposition of the equation of state, and specifies the scale factor as 
\be
\label{sf2} a(t) = a_s \left[\delta + \left(1 - \delta \right) y^m -
\delta \left( 1 - y \right)^n \right] \hspace{0.1cm} y \equiv
\frac{t}{t_s}~,
\ee
where $a_s \equiv a(t_s)=$ const. and $\delta, m, n =$ const.,
%\bea
%\label{dota}
%\dot{a} &=& a_s \left[ \frac{m}{t_s} \left(1 - \delta \right)y^{m-1} + \delta\frac{n}{t_s}\left( 1 - y \right)^{n-1} \right]~,\\
%\label{ddota}
%\ddot{a} &=& \frac{a_s}{t_s^2} \left[ m\left(m-1\right) \left(1 -
%    \delta \right) y^{m-2} \right. \nonumber \\
%    &-& \left. \delta n \left(n-1 \right) \left(1-y \right)^{n-2} \right]~,
%\eea
then one gets (apart from a Big-Bang at $t=0$) a new type of singularity at $t=t_s$ (provided $1 < n < 2$) which was christened a Sudden Future Singularity (SFS). Such a singularity is a singularity of pressure $p$ (or $\ddot{a}$) and leads merely to DEC violation. The standard ''Friedmann limit'' is easily obtained by taking the ``nonstandardicity'' parameter $\delta \to 0$ in (\ref{sf2}). In fact, at SFS we have:
 \bea
a = {\rm const.}, \hspace{0.5cm} \dot{a} = {\rm const.} \hspace{0.5cm} \varrho = {\rm const.}
\nonumber \\
 \ddot{a} \to - \infty \hspace{0.5cm} p \to \infty \hspace{0.5cm} {\rm
   for } \hspace{0.5cm} t \to t_s~.
\eea
It is interesting that the Schwarzschild horizon at $r=r_g$ has a singular metric, while the curvature invariants are regular there. On the ohter hand, an SFS at $t=t_s$ has a regular metric, while curvature invariants diverge. 

The matter related to SFS may serve as dark energy, especially if they are quite close in the near future. For example, an SFS may even appear in 8.7 Myr with no contradiction to a bare supernovae data. It can be fitted to a combined SnIa, CMB and BAO data, but at the expense of admitting on the approach to a Big-Bang a fluid which is not exactly dust (m=0.66), but has slightly negative pressure (m = 0.73, w = -0.09) \citep{PRD12}. A more general class of singularities known as Generalized Sudden Future Singularities (GSFS) which do not violate any of the energy conditions are also possible \citep{GSFS}.

There is yet the whole class of non-Big-Bang singularities \citep{nojiri,aip10} (Finite Scale Factor, Big Separation, $w$-singularities \citep{wsing}, Little-Rip \citep{LRip}, Pseudo-Rip \citep{PRip} and their various versions like Big-Boost and Big-Brake (belonging to an SFS class \citep{BBrake}), Big-Freeze (belonging to an FSF class \citep{BFreeze}), generalized Big-Separation \citep{yurov} and generalized $w$-singularities \citep{yurov}). Most of them can be described using one unified scale factor \citep{JCAP13} reading as
\be \label{newscalef} a(t) = a_s \left( \frac{t}{t_s} \right)^m \exp{\left( 1 - \frac{t}{t_s} \right)^n} ~, \ee
with the appropriate choice of constants $t_s, a_s, m, n$. In fact, from (\ref{newscalef}) we can see that for $0 < m < 2/3$ we deal with a BB singularity and $a \to 0$, $\varrho \to \infty$, $p \to \infty$ at $t \to 0$. For $m < 0$ we have a BR singularity with $a \to \infty$, $\varrho \to \infty$, $p \to \infty$ at $t=0$. An SFS appears for $1<n<2$ at $t=t_s$ ($a=a_s$, $\varrho=$ const., $p \to \infty$), and an FSF appears for $0<n<1$ at $t=t_s$ ($a=a_s$, $\varrho \to \infty$, $p \to \infty$).

In order to classify the strength of standard and exotic singularities some definitions have been proposed. According to \citet{tipler} a singularity is weak if a double integral
\be
\int_0^{\tau} d\tau' \int_0^{\tau'} d\tau'' R_{ab}u^a u^b
\ee
does not diverge on the approach to a singularity at $\tau = \tau_s$ ($\tau$ is the proper time), while according to \citet{krolak} a singularity is weak if a single integral
\be
\int_0^{\tau} d\tau' R_{ab}u^a u^b
\ee
does not diverge on the approach to a singularity at $\tau = \tau_s$. Otherwise, a singularity is strong. From now on, T will stand for the definition of Tipler, and K will stand for the definition of Kr\'olak. It is interesting that both point particles and even extended objects may not feel weak singularities and can pass through them \citep{leonardo,PRD06}.

Classification of exotic singularities was first given by \citet{nojiri} and further developed by \citet{aip10}. Their current classification is attempted in Table \ref{classif}.

\begin{table*}
\caption{Classification of singularities in FRW cosmology}
\label{classif}
\begin{center}
\begin{tabular}{lcccccccccc}
\hline
\\
Type & Name & t sing. & a(t$_s$) & $\varrho(t_s)$ & p(t$_s$) & $\dot{p}(t_s)$ etc.  & w(t$_s$) & T & K\\
\hline
\\
0  & Big-Bang (BB) & $ 0 $ & $ 0$ & $\infty$ & $\infty$ &$\infty$& finite & strong & strong\\
I  & Big-Rip (BR) & $t_s $ &$\infty$ & $\infty$ & $\infty$ &$\infty$& finite & strong & strong\\
I$_l$  & Little-Rip (LR) & $\infty$ &$\infty$ & $\infty$ & $\infty$ & $\infty$ & finite & strong & strong\\
I$_p$  & Pseudo-Rip (PR) & $\infty$ &$\infty$ & finite & finite & finite & finite & weak & weak\\
II  & Sudden Future (SFS) & $t_s$ & $a_s$ & $\varrho_s$ & $\infty$ & $\infty$ & finite & weak & weak\\
II$_g$  &Gen. Sudden Future (GSFS) & $t_s$ &$a_s$ & $\varrho_s$ & $p_s$ &$\infty$& finite & weak & weak\\
III  & Finite  Scale  Factor (FSF) & $t_s$ &$a_s$ & $\infty$ & $\infty$ &$\infty$& finite& weak & strong\\
IV & Big-Separation  (BS) & $t_s$ &$a_s$ & $0$ & $0$ &$\infty$& $\infty$ & weak & weak\\
V & w-singularity (w) & $t_s$ &$a_s$ & $0$ & $0$ &$0$& $\infty$& weak & weak
\\
\hline
\end{tabular}
\end{center}
\end{table*}

\section{Varying constants theories.}

First fully quantitative framework which allowed for variability of the fundamental constant was Brans-Dicke scalar-tensor gravity. The gravitational
constant $G$ in such a theory is associated with an average gravitational potential (scalar field) $\Phi$ surrounding a given particle:
$<\Phi> = GM/ (c/H_0) \propto 1/G = 1.35 \times 10^{28} g/cm$. The scalar field $\Phi$ gives the strength of gravity
\be
G = \frac{1}{16 \pi \Phi}~,
\ee
which changes Einstein-Hilbert action into Brans-Dicke action
\be
S = \int d^4 x \sqrt{-g} \left( \Phi R - \frac{\omega}{\Phi} \partial_{\mu} \Phi \partial^{\mu} \Phi + L_m \right)~,
\ee
and further relates to low-energy-effective superstring action when $\omega = - 1$. In superstring theory, the string coupling constant $g_s = \exp{(\phi/2)}$ changes in time with $\phi$ being the dilaton, and $\Phi = \exp{(-\phi)}$.

Another framework is given by varying speed of light theories (VSL). In \citet{VSL} model (AM) the speed of light is replaced by a scalar field
\be
c^4 = \psi(x^{\mu})~~,
\ee
leading to the action
\be
S = \int d^4 x \sqrt{-g} \left[ \frac{\psi R }{16 \pi G} + L_m  + L_{\psi} \right]~~.
\ee
AM model breaks Lorentz invariance (relativity principle and light principle). There is a preferred frame (called a cosmological or a CMB frame)
in which the field is minimally coupled to gravity. This model solves basic problems of standard cosmology such as the horizon problem and the flatness problem.
One of the ans\"atze is that $\rho = \rho_0 a^{-3(w+1)}$, $c(t) = c_0 a^n$ which solves the above two problems if $n \leq -(1/2)(3w+1)$ \citep{VSL}.
Another version of VSL is Magueijo covariant (conformally) and locally invariant model \citep{Magueijo2000}:
\be
\psi = \ln{\left( \frac{c}{c_0} \right)} \hspace{0.5cm} {\rm or} \hspace{0.5cm} c = c_0 e^{\psi}~,
\ee
with the action
\be
S = \int d^4 x \sqrt{-g} \left[ \frac{c_0^4 e^{\alpha \psi} (R + L_{\psi})}{16 \pi G} + e^{\beta \psi} L_m  \right]~,
\ee
and
\be
L_{\psi} = \kappa(\psi) \nabla_{\mu}\psi \nabla^{\mu} \psi~.
\ee
There is an extra condition that $\alpha - \beta = 4$ with interesting subcases: a) $\alpha = 4; \beta = 0$, giving Brans-Dicke theory with $\phi_{BD} = e^{4\psi}/G$ and $\kappa(\psi) = 16 \omega_{BD}(\phi_{BD})$; b) $\alpha = 0; \beta = -4$, called a minimal VSL theory.

Yet another framework is varying fine structure constant $\alpha$ theory (or varying charge $e = e_0 \epsilon(x^{\mu})$ theory \citep{webb1999}
\bea
S &=& \int d^4 x \sqrt{-g} \left( \psi R - \frac{\omega}{2} \partial_{\mu} \psi \partial^{\mu} \psi \nonumber \right. \\
&-& \left. \frac{1}{4} f_{\mu\nu}f^{\mu\nu} e^{-2\psi} + L_m \right)
\eea
in which the scalar field is associated with electric charge $\psi = \ln{\epsilon}$ and $f_{\mu\nu} = \epsilon F_{\mu\nu}$.
This model can be related to the VSL theories due to the definition of the fine structure constant $\alpha(t) = e^2/[\hbar c(t)]$.
If one assumes the linear expansion of $e^{\psi} = 1 - 8 \pi G \zeta (\psi - \psi_0) = 1 - \Delta \alpha/\alpha$ with the constraint on the local equivalence principle violence $\mid \zeta \mid \leq 10^{-3}$, then the relation to dark energy density parameter $\Omega_{\psi}$ is 
\be
w+1 = \frac{(8\pi G \frac{d\psi}{d\ln{a}})^2}{\Omega_{\psi}}~,
\ee
which can be tested (while mimicking the dark energy) by spectrograph CODEX (COsmic Dynamics EXplorer) -- a device attached to a planned E-ELT (European Extremely Large Telescope) measuring the redshift drift effect (or Sandage-Loeb effect - \citet{sandage,loeb}) for $2 < z <5$ \citep{martins}.

\section{Varying constant versus cosmic singularities.}

It has been shown that quantum effects \citep{houndjo} may change the strength of exotic singularities (e.g. an SFS to become an FSF). As it was already mentioned, varying constants cosmologies have been applied to solve standard cosmology problems as well. Our idea is to apply them to solve the singularity problem in cosmology. We can also ask if varying constants theories can soften/strengthen the standard and exotic singularities?

We consider the Friedmann universes in varying speed of light (VSL) theories and varying gravitational constant G theories as follows
($\varrho$ - mass density; $\varepsilon = \varrho c^2(t)$ - energy density in $Jm^{-3} = Nm^{-2} = kgm^{-1}s^{-2}$)
\bea \label{rho} \varrho(t) &=& \frac{3}{8\pi G(t)}
\left(\frac{\dot{a}^2}{a^2} + \frac{kc^2(t)}{a^2}
\right)~,\\
\label{p} p(t) &=& - \frac{c^2(t)}{8\pi G(t)} \left(2 \frac{\ddot{a}}{a} + \frac{\dot{a}^2}{a^2} + \frac{kc^2(t)}{a^2} \right)~,
\eea
with the source terms in the energy-momentum ``conservation law'' due to varying $c$ and $G$:
\bea
\label{conser}
\dot{\varrho}(t) + 3 \frac{\dot{a}}{a} \left(\varrho(t) + \frac{p(t)}{c^2(t)} \right) = - \varrho(t) \frac{\dot{G}(t)}{G(t)}
+ 3 \frac{kc(t)\dot{c}(t)}{4\pi Ga^2}~.\nonumber
\eea
For flat $k=0$ universes we have
\bea
\label{rhot}
\varrho(t) &=& \frac{3}{8\pi G(t)} \left[\frac{m}{t} - \frac{n}{t_s} \left( 1 - \frac{t}{t_s} \right)^{n-1} \right]^2~,\\
\label{pt}
p(t) &=& - \frac{c^2(t)}{8 \pi G(t)} \left[ \frac{m(3m-2)}{t^2} - 6 \frac{mn}{tt_s} \left( 1 - \frac{t}{t_s} \right)^{n-1} \right. \nonumber \\
&+& \left. 3 \frac{n^2}{t^2_s} \left( 1 - \frac{t}{t_s} \right)^{2(n-1)} \right. \\
&+& \left. 2 \frac{n(n-1)}{t^2_s} \left( 1 - \frac{t}{t_s} \right)^{n-2} \right]~. \nonumber
\eea
One bears in mind the scale factor (\ref{newscalef}), the mass density (\ref{rhot}), and the pressure (\ref{pt}).

\subsection{Regularizing a Big-Bang singularity by varying $G$}

If $G(t) \propto 1/(t^2)$, which is a faster decrease than in Dirac's Large Number Hypothesis (LNH) $G \propto 1/t$, and influences less the temperature of the Earth constraint \citep{teller}, then both divergence in $\varrho$ and $p$ are removed, though at the expense of having the "singularity" of strong gravitational coupling $G \to \infty$ at $t \to 0$. In the Dirac's case, only the singularity in $\varrho$ can be removed.

\subsection{Regularizing an SFS singularity by varying $c$}

If
\be
\label{regc}
c(t) = c_0 \left( 1 - \frac{t}{t_s} \right)^{\frac{p}{2}}~,
\ee
then
\bea
\label{pregc}
&&p(t) = - \frac{c^2_0}{8\pi G} \left[ \frac{m(3m-2)}{t^2}\left( 1 - \frac{t}{t_s} \right)^p \right.  \\
&-& \left. 6 \frac{mn}{tt_s} \left( 1 - \frac{t}{t_s} \right)^{p+n-1}
+ 3 \frac{n^2}{t^2_s} \left( 1 - \frac{t}{t_s} \right)^{p + 2n -2} \nonumber \right. \\
&+& \left. 2 \frac{n(n-1)}{t^2_s} \left( 1 - \frac{t}{t_s} \right)^{p+n-2} \right]. \nonumber
\eea
and the singularity of pressure is regularized provided $p > 2 - n, (1<n<2)$.

Physical consequence of such a choice of regularization is that light eventually stops at singularity: $c(t_s) = 0$. Same happens in loop quantum cosmology (LQC), where we deal with the anti-newtonian limit $c = c_0 \sqrt{1- \varrho/\varrho_c} \to 0$ for $\varrho \to \varrho_c$ with $\varrho_c$ being the critical density \citep{cailleteau}. The low-energy limit $\varrho \ll \varrho_0$ gives the standard case $c \to c_0 =$ const. It also appears naturally in \citet{Magueijo2001} model, in which black holes are not reachable since the light stops at the horizon (despite they possess Schwarzschild singularity). Both $c = 0$ and $c = \infty$ options are possible in Magueijo model.

\subsection{No way of regularizing a $w$-singularity by varying $c$}

In the limit $m \to 0$ of (\ref{newscalef}) we have an exotic singularity scale factor given by $a(t) = a_s \exp{(1-t/t_s)}$, and from (\ref{rhot}) and (\ref{pt}) we get
\bea
\label{exrhot}
\varrho_{ex}(t) &=& \frac{3}{8\pi G(t)} \frac{n^2}{t_s^2} \left( 1 - \frac{t}{t_s} \right)^{2(n-1)}~,\\
\label{expt}
p_{ex}(t) &=& - \frac{c^2(t)}{8 \pi G(t)} \left[ 3 \frac{n^2}{t^2_s} \left( 1 - \frac{t}{t_s} \right)^{2(n-1)} \right. \nonumber \\
&+& \left. 2 \frac{n(n-1)}{t^2_s} \left( 1 - \frac{t}{t_s} \right)^{n-2} \right]~
\eea
so that
\be
\label{exw}
w_{ex}(t) = \frac{p_{ex}(t)}{\varepsilon_{ex}(t)} = - \left[ 1 + \frac{2}{3} \frac{n-1}{n} \frac{1}{\left(1-\frac{t}{t_s}\right)^n} \right]~,
\ee
which is a $w$-singularity for $n>2$ ($p = \varrho = 0$, $w_{ex} \to \infty$) \citep{wsing}. Its regularization by varying $c(t)$ is impossible since there is no $c$-dependence here.

\subsection{Regularizing an SFS singularity by varying $G$}

If we assume that
\be
\label{regG}
G(t) = G_0 \left( 1 - \frac{t}{t_s} \right)^{-r}~~,
\ee
($r=$ const., $G_0=$ const.) which changes (\ref{rhot}) and (\ref{pt}) to
\bea
\label{rhoG}
\varrho(t) &=& \frac{3}{8\pi G_0} \left[\frac{m^2}{t^2} \left( 1 - \frac{t}{t_s} \right)^r - \frac{2mn}{tt_s} \left( 1 - \frac{t}{t_s} \right)^{r+n-1} \right. \nonumber \\
&+& \left. \frac{n^2}{t^2_s} \left( 1 - \frac{t}{t_s} \right)^{r+2n-2} \right]~,\\
\label{pG}
p(t) &=& - \frac{c^2}{8\pi G_0} \left[ \frac{m(3m-2)}{t^2}\left( 1 - \frac{t}{t_s} \right)^r \right. \nonumber \\
&-& \left. 6 \frac{mn}{tt_s} \left( 1 - \frac{t}{t_s} \right)^{r+n-1} + 3 \frac{n^2}{t^2_s} \left( 1 - \frac{t}{t_s} \right)^{r + 2n -2} \right. \nonumber \\
&+& \left. 2 \frac{n(n-1)}{t^2_s} \left( 1 - \frac{t}{t_s} \right)^{r+n-2} \right]~.
\eea
From (\ref{rhoG}) and (\ref{pG}), it follows that an SFS singularity $(1<n<2)$ is regularized by varying gravitational constant when
$r> 2-n$, and an FSF singularity $(0<1<n)$ is regularized when $r> 1-n$. On the other hand, assuming that we have an SFS singularity and that
$-1<r<0$, we get that varying $G$ may change an SFS singularity onto a stronger FSF singularity when $0 < r+n < 1$.

\section{Conclusions}

Our proposal was to investigate how the standard and exotic FRW singularities are influenced by varying physical constants. In particular, we were looking for the answer if it was possible to "regularize" (remove infinities) or change these singularities and what were the physical consequences of such an action, because what we faced was often the new singularity in a physical constant/field which acted to remove/change the type of singularity.

We have shown that in order to regularize an SFS or an FSF singularity by varying $c(t)$, the light should slow and eventually stop propagating at a singularity. Similar effects were found in loop quantum cosmology (LQC) as well as in VSL theory for Schwarzschild horizon, where the speed of light was going either zero or to infinity at $r=r_s$. An observer could not reach this surface even in his finite proper time.

In order to regularize an SFS or an FSF by varying gravitational constant $G(t)$, the strength of gravity has to become infinite at singularity. It seems reasonable because of the requirement to overcome an infinite (anti-)tidal forces. On the other hand, it makes another singularity - a singularity of strong coupling for a physical field such as $G \propto 1/\Phi$. Such problems were already dealt with in superstring and brane cosmology where both the curvature singularity and a strong coupling singularity appeared (and requires special choice of coupling, or the application of quantum corrections).

\begin{acknowledgements}
This project was financed by the National Science Center Grant DEC-2012/06/A/ST2/00395.
\end{acknowledgements}

\bibliographystyle{aa}

\end{document}